# X (3872) as a Universal and Composite State

B. Kabirimanesh*, H. Mehraban**

Physics Department, Semnan University
P.O. Box 35195-363, Semnan, Iran

**Abstract**

X (3872) because of its mass that is just below $D^0 \bar{D}^{*0}$ threshold and its quantum number $J^{PC} = 1^{++}$ can be considered as a weakly bound molecule. It can be regarded as non-relativistic system depicting universality and compositeness. Physical observables are called universal if they are insensitive to the range and other details of short-range interaction. Weinberg introduced compositeness criterion to classify a two-body composite system as the sum, with different weights, of a compact core and a two distinct parts.

## 1. Introduction

Hadron spectroscopy opened a new era In hadron physics called exotics. Exotics cannot be explained by conventional view of $q\bar{q}$ meson and $qqq$ baryon. $QCD$ does not prohibit hadrons made of two quarks and two antiquarks (tetraquarks), four quarks and one antiquark (pentaquark), six quarks (hexaquarks) and other combinations of quarks and gluons called hybrids and configuration of gluons or glueballs [1-7].

Various theoretical models was proposed to explain the rich phenomena of hadronic spectroscopy. A compact tetraquark is formed by diquark and antidiquark [8,9]. A hybrid is a quark antiquark meson with exited gluons in addition to the rotational and the radial motion of ordinary quarks [10,11]. Hadroquarkonium is a compact heavy quark antiquark core surrounded by light quark cloud and the QCD analogue of the Vander Waals force holding them together [12,13]. Hadronic molecule is a compound object consisted of two or more hadrons. The hadronic molecule is founded on the deuteron which can be explained as shallow bound state of proton and neutron [14]. Hadronic molecules are loosely bound, it means that the binding energies are small with respect to QCD scale $\Lambda_{QCD}$ which is several hundred MeV.

___

*E-mail: Babakkabirimanesh@semnan.ac.ir
**E-mail: hmehraban@semnan.ac.ir

In 2003 the Belle collaboration discovered a new structure producing by the decay $B^{\pm} \to K^{\pm} X$ and observed through decay mode $J/\psi\, \pi^+ \pi^-$ called X (3872) [15] and later confirmed by CDF [16]. This state had predicted 10 years earlier by Tornqvist [17].

The average mass of X (3872) can be estimated as [18]

$$M_X = 3871.69 \pm 0.17 MeV.$$

For the width only upper limit exist [19]

$$\Gamma_X < 1.2 MeV,$$

and it is found that the two-pion and three –pion branching ratios were of the same order of magnitude [20]

$$\frac{Br[X \to J/\psi\, \pi^+ \pi^- \pi^0]}{Br[X \to J/\psi\, \pi^+ \pi^-]} = 1.0 \pm 0.4 \pm 0.3.$$

Researches have shown that that the dipion in the $J/\psi\, \pi^+ \pi^-$ mode comes from isovector $\rho$ meson while the tripion in the $J/\psi\, \pi^+ \pi^- \pi^0$ mode comes from the isoscalar $\omega$ meson so this fraction shows strong isospin violation.

The analogue of X (3872) in the bottom sector has not been yet found. The search for it has done by CMS Collaboration but no signal was detected in $\Upsilon \pi^+ \pi^-$ channel [21-23].

The low energy Few-body observables for non-relativistic particles with short range interaction and large scattering length have universal properties that are not tied to the detail of mechanism that creates large scattering length. This phenomena is called low energy universality. For $a > 0$ the elementary prediction is that there is a 2-body shallow bound state whose binding energy for large $a$ approaches [24]

$$B \to \frac{1}{2\mu a^2}, \qquad \mu = \frac{m_1 m_2}{m_1 + m_2}, \qquad (1)$$

here $\mu$ is the reduced mass of two parts and $a$ denotes scattering length. If binding energy of X (3872) is 0.5MeV its scattering length would be 6.3fm that is much larger than natural length scale $\frac{1}{m_\pi} = 1.5\ fm.$

There is a universal prediction for the $D^0 \bar{D}^{*0}$ or $\bar{D}^0 D^{*0}$ wave function as [25]

$$\psi(r) \longrightarrow (2\pi a)^{-1/2} \frac{\exp(-r/a)}{r} . \tag{2}$$

The scaling limit is an important concept that can be defined by describing phase shift $\delta_0(\boldsymbol{k})$ for two body scattering. The phase shift has a simple form as [26,27]

$$k \cot \delta_0(\boldsymbol{k}) = -\frac{1}{a} . \tag{3}$$

The Weinberg compositeness criterion states that for the physical wave function of a bound state [28-30]

$$|\psi\rangle = \begin{pmatrix} \lambda |\psi_0\rangle \\ \chi(\boldsymbol{k}) |h_1 h_2\rangle \end{pmatrix}, \tag{4}$$

here $|\psi_0\rangle$ represents the compact part of the state and $|h_1 h_2\rangle$ is two hadron part and $R_{conf} < 1\,fm$. If $\gamma$ denotes binding momentum the radius of confinement is more compact than $R \sim 1/\gamma$ which represents the characteristic size of shallow bound state. $\chi(\boldsymbol{k})$ is the wave function of the two hadron part and $\boldsymbol{k}$ represents the relative momentum of two particles. $\lambda^2$ signifies the probability of compact part of wave function in the physical state which is analogous to wave function renormalization constant $Z$ in quantum field theory [27]. The system could be described by the Schrödinger equation

$$\hat{H}|\psi\rangle = E|\psi\rangle , \quad \hat{H} = \begin{pmatrix} \hat{H}_{cc} & \hat{V} \\ \hat{V} & \hat{H}^0_{hh} \end{pmatrix}. \tag{5}$$

For two hadron Hamiltonian the kinetic energy is given by $\hat{H}^0_{hh} = k^2/(2\mu)$ and the transition form factor is

$$\langle \psi_0 | \hat{V} | h_1 h_2 \rangle = f(\boldsymbol{k}), \tag{6}$$

where

$$\chi(\boldsymbol{k}) = \lambda \frac{f(\boldsymbol{k})}{E - k^2/(2\mu)}, \tag{7}$$

and for scattering length and effective range we have

$$a = -2\frac{1-\lambda^2}{2-\lambda^2}(\frac{1}{\gamma}) + O(\frac{1}{\beta}), \tag{8}$$

$$r = -\frac{\lambda^2}{1-\lambda^2}(\frac{1}{\gamma}) + O(\frac{1}{\beta}), \tag{9}$$

where $\beta$ is the inverse range of force. For pure molecule we have $\lambda^2 = 0$ and for compact state $\lambda^2 = 1$. Writing it in terms of renormalization constant we have [31]

$$a = 2(\frac{1-Z}{2-Z})R + O(\frac{1}{m}) \quad , \quad r = -(\frac{Z}{1-Z})R + O(\frac{1}{m}) \quad , \quad R = \sqrt{\frac{1}{2\mu B}}, \tag{10}$$

where $R$ is characteristic size of state and $m$ represents corrections due to momentum scale.

The residue of a bound-state pole located close to the threshold could be shown as [27]

$$\frac{g_{eff}^2}{4\pi} = 4M^2(\frac{\gamma}{\mu})(1-\lambda^2), \tag{11}$$

where $g_{eff}$ is effective coupling constant of the bound system and it is maximum for pure molecule $\lambda^2 = 0$. Weinberg criterion was expanded for resonances and are discussed for couple channels [32-34]. It is crucial to note that in order to identify hadronic molecules, those observables are useful that are sensitive to effective coupling expressed in Eq.(11).

In the coupled channel scheme in which a resonance is a mixture of a bare elementary state and multiple two body components by using the spectral density concept we have

$$\int_0^\infty w(E)dE = 1 - \sum_\alpha Z_\alpha, \tag{12}$$

where $w(E)$ is the spectral density which gives the probability to find bare state in the continuum wave function and $Z_\alpha$ gives the probability to find a bare state in the wave function of the $\alpha$-th bound state [35,36].

For unstable states the generalization of weak binding relation is obtained by adding a two-body decay channel in the effective field theory. The threshold energy of the decay channel is set at $E = -v$ with $v > 0$ measured from threshold of channel 1.

$$|h\rangle = \sqrt{X}|Channel1\rangle + \sqrt{1-X}|others\rangle, \tag{13}$$

here $|h\rangle$ represents the quasi-bound state of the wave function and X shows the composite component of channel 1.

The weak- binding relation for quasi-bound state $|h\rangle$ can be written as

$$a = R\left\{\frac{2X}{1+X} + O\left(\left|\frac{R_{typ}}{R}\right|\right) + O\left(\left|\frac{l}{R}\right|^3\right)\right\},\qquad(14)$$

with

$$R = \sqrt{\frac{1}{-2\mu E_h}}, \qquad l = \sqrt{\frac{1}{2\mu\nu}}, \qquad (15)$$

here $a$ is the scattring length of channel 1 and $l$ is the length scale associated with the threshold energy difference $\nu$. The first and second term take the same form as in the weak binding relation for the bound state when $\left|\frac{l}{R}\right|^3 \ll 1$ it can be neglected [37].

**2. The X (3872)**

Wave function of the X (3872) should be expressed in fock decomposition [38]

$$\psi_X = a_0\psi_0 + \sum_i a_i\psi_i, \qquad (16)$$

here $\psi_0$ is the state of neutral $D$ mesons $(D^0\bar{D}^{*0} + \bar{D}^0 D^{*0})/\sqrt{2}$ and $\psi_i$ to other hadronic states.

The result found by reference [39] is

$$|X\rangle = 0.237|c\bar{c}\rangle - 0.944|D^0\bar{D}^{*0}\rangle - 0.228|D^+D^{*-}\rangle$$
$$= 0.237|c\bar{c}\rangle - 0.829|D\bar{D}^*; I=0\rangle - 0.506|D\bar{D}^*; I=1\rangle.$$

. Despite of large phase space of the X (3872) for decay to $Q\bar{Q}$ and pions it has narrow width that means the wave function have a very small overlap with the corresponding quarkonia so it better it to interpret it as molecule [40].

Bound state of the X (3872) with no direct interaction in the mesonic channel gives [41]

$$\frac{1}{Z} - 1 \approx \frac{4\pi^2\mu^2 f_0^2}{\sqrt{2\mu B}}, \qquad (17)$$

where $f_0$ is the bare coupling constant between channels, by decreasing the binding energy $Z$ also decreases and it means a larger molecular component.

The features of the X (3872) can be explained in a universal EFT with contact interaction but there is no Efimov effect in the system. [24,42,43]

Large scattering length is also studied in Feshbach resonance. Feshbach resonance first appeared in literature to explain the narrow resonances observed in the total cross section for a neutron scattering of nucleus [44]. The idea of dressed molecule which is the important idea of many-body Feshbach physics is superposition of the closed channel bound state or pure molecule and the set of states that are in open channel channel continuum [25,45].

$$|dressed\rangle = \sqrt{Z}|closed\rangle + \sqrt{1-Z}|open\rangle.$$

**Table 1** The scattering length $a$ and effective range $r$ are calculated in *fm* for the *X(3872)* by using Eq. (10) for different values of $Z$. The masses are taken from Ref. [18]. We take $B = 0.18\,MeV$.

| $Z$ | $Z = 0$ | $Z = 0.1$ | $Z = 0.3$ | $Z = 0.5$ | $Z = 0.7$ | $Z = 0.9$ | $Z = 1$ |
|---|---|---|---|---|---|---|---|
| $a$ | 12.40 | 11.80 | 10.40 | 8.80 | 6.40 | 2.60 | 1.40 |
| $r$ | 1.40 | 0.20 | -3.40 | -9.60 | -24.20 | -97.60 | - |

In table 1. we use Eq.(10) and we take binding energy of the X (3872) 0.18 MeV. In order to calculate the scattering length and effective range the value of $O(\frac{1}{\beta})$ where $\beta$ is the inverse range of force is estimated as $\frac{1}{m_\pi}$, where $m_\pi$ is the mass of $\pi$ meson. With this value of the binding energy we find large positive value for $a$ but negative value for $r$ for $Z \geq 0.3$. Thus if the X (3872) is a dominant molecular structure $Z < 0.3$.

## 3. Conclusion

There exist different models for interpreting the X. The X has been observed in various decay modes more than other heavy exotics. But there is no common view about its nature. X can be created in any reaction that can generate its ingredients $(D^0 \bar{D}^{*0} + \bar{D}^0 D^{*0})/\sqrt{2}$, $D^0 \bar{D}^{*0}$ and $\bar{D}^0 D^{*0}$.

Since the mass X is close to s-wave threshold and its quantum number is similar to analogous meson pair and has a narrow width, these characteristics motivated some physicists to take it as threshold cusps [46,47].

The analogy between X and deuteron is motivating, but deuteron is observed directly by studying its stopping power, but the X cannot be observed directly. The X(3872) and deuteron have been created in high energy pp collisions. But the reason how such loosely bound heavy states be created had not been yet uncovered.

Because accurate mass of the X(3872) is important to understand its nature, new proposals based on triangle singularities $ee^+ \to X\gamma$ have been suggested [48,49].

Regarding to heavy quark spin symmetry [50] there should be more hadronic molecules degenerate with the X(3872) [51-53].

Lattice calculation for the X(3872) are not so successful. For the binding energy of the X(3872) the uncertainty is comparable to its central value [54,55]. The limitations related to lattice calculation can be recovered by the new numerical methods and increasing the computer power.

Currently we don't know on which sheet the pole of the X(3872) is located. It can be a shallow bound state which means which means it has a pole on a physical sheet or virtual state which means it has a pole on unphysical sheet [56,57].

Weinberg approach that whether or not the deuteron is composite or elementary gives insight s to other hadronic resonances like X(3872).

The X(3872) could be a loosely bound charm meson molecule comparable in size to the largest nuclei. Because of very small binding energy, it makes nice area for implication of effective field theory.

More measurements and detection of a large number of new decays of the X(3872) could reveal its internal structure and production mechanism.